\documentstyle{laa} 
\input{psfig}
\begin{document}

\thesaurus{03(
	      11.04.1; %Galaxies: distances, redshifts 
	      11.04.2; %Galaxies: dwarf 
	      11.06.2; %Galaxies: fundamental parameters 
	      11.07.1; %Galaxies: general 
	      }  

\title{HI properties of nearby galaxies from a volume-limited sample 
	   \thanks{The Appendix is available only in electronic form at
	   the CDS via anonymous ftp(139.79.128.5)}}

\author{I.D.Karachentsev,  D.I.Makarov \inst{1} \and W.K.Huchtmeier \inst{2}}

\offprints{W. K. Huchtmeier; \it email: huchtmeier@mpifr-mpg.de} 

\institute{Special Astrophysical Observatory, Russian Academy of Sciences,
		    N.Arkhyz, KChR, 357147, Russia
\and Max-Planck-Institut f\"{u}r Radioastronomie, Auf dem Hugel 69,
		    D-53121 Bonn, Germany}

\maketitle

\begin{abstract}
   We consider global HI and optical properties of about three hundred
nearby galaxies with  V$_0 < 500$ km/s. The majority of them have individual
photometric distance estimates. The galaxy sample parameters show some
known and some new correlations implying a meaningful dynamic explanation:
1) In the whole range of diameters, 1 --- 40 Kpc, the galaxy standard
diameter and rotational velocity follows a nearly linear Tully-Fisher
relation, $\lg A_{25}\propto(0.99\pm0.06)\lg V_m.$\hspace{0.2cm} 2) The HI mass-to-luminosity
ratio and the HI mass-to-"total" mass (inside the standard optical diameter)
 ratio increase systematically from giant
galaxies towards dwarfs, reaching  maximum values $5 M_{\sun}/L_{\sun}$ and 3,
respectively. 3) For all the Local Volume galaxies their total mass-to-
luminosity ratio lies within a range of [0.2--16] $M_{\sun}/L_{\sun}$ with a median
of $3.0 M_{\sun}/L_{\sun}$. The $M_{25}/L$ ratio decreases slightly from giant
 towards dwarf galaxies. 4) The $M_{HI}/L$ and $M_{25}/L$ ratios for the
sample galaxies correlate with their mean optical surface brightness,
which may be caused by star formation activity in the galaxies.
5) The $M_{HI}/L$ and $M_{25}/L$ ratios are practically independent of the local
mass density of surrounding galaxies within the range of densities of about
six orders of magnitude. 6) For the LV galaxies their HI mass and angular momentum
follow a nearly linear relation: $\lg M_{HI}\propto(0.99\pm0.04)\lg(V_m\cdot A_{25})$,
expected for rotating gaseous disks being near the threshold of gravitational
instability, favourable for active star formation.
\end{abstract}

\section{Introduction}
  The acquision of global parameters of galaxies from observations in the
neutral hydrogen line (Roberts, 1969) revealed a new approach in investigating
the evolution of gaseous and stellar subsystems of galaxies. Over the past
30 years the number of galaxies with measured HI flux and line width has been
increased by two orders, reaching $N\sim10^4$. Numerous publications were
devoted to study relations between global HI and optical characteristics
of galaxies. A vast comparison of HI data for about 6 000 galaxies
was undertaken by Roberts \& Haynes (1994). However, their sample, as many
others, are limited by apparent magnitude or angular diameter of galaxies,
but not by distance, which leads to a strong observational selection
against intrinsically small objects.

  For the analysis of global parameters of galaxies in a volume- limited
sample Huchtmeier \& Richter (1988) (=HR) used the list of galaxies from
Kraan-Korteweg \& Tammann (1979) (=KKT) with corrected radial velocities,
$V_0 < 500$ km/s. The sample studied by HR contained 146 galaxies mainly
of low luminosity. For the past decade the initial KKT sample has been
increased almost two times in number due to the mass surveys of redshifts
of galaxies from the known catalogues, revealing  new nearby galaxies in
the Milky Way "Zone of Avoidance", as well as special searches for dwarf
galaxies in nearby groups. The increasing numbers of galaxies in the LV is 
mainly due to many new dwarf galaxies. It should be emphisized that the numerical growth
of the KKT sample was accompanied by improving its data quality. For
many faint galaxies their apparent magnitudes, angular diameters and types
were specified. However, even at present one may find cases of significant
discrepancy in the published data.

  It should be particularly emphasized that for galaxies with $V_0 < 500$ km/s
their corrected radial velocity is a rather unreliable distance indicator
because of observed deviations from the perfect cosmic Hubble flow,
V = HD (Karachentsev \& Makarov, 1996). That is why
measurements of individual photometric distances for $\sim150$
Local Volume galaxies made by different authors in the 90-ies allowed
masses and other global parameters of nearby galaxies to be determined with
a higher accuracy (see references in Karachentsev (1994) and Karachentsev \&
Tikhonov (1994)). The mentioned circumstances give us  grounds
to consider once again the relations between the global parameters of nearby
galaxies based on a more complete and homogeneous sample of observational
data.

\section{Initial data}
  As the main source of data on optical and HI properties of galaxies, we
used the last version (LEDA) of the Principal Galaxy Catalogue (Paturel et
al., 1992). For many galaxies their apparent magnitudes, angular diameters
and morphological types were revised using  large scale CCD frames
obtained with the SAO 6-meter telescope and the 2.6-meter Nordic telescope.
>From the list of galaxies with $V_0 < 500$ km/s  we excluded about 100 objects
in the central region of the Virgo cluster and also more than 50 fictitious
"nearby galaxies" (i.e. globular clusters, planetary nebulae or cases when the
radial velocity is inherent in a bright star projected on a distant galaxy).
Our sample (the Local Volume sample = LV) is updated by new objects from
the list of Karachentseva \& Karachentsev (1998) and other recent sources.
Particularly, there were taken into account the results of HI searches for
nearby dwarf galaxies made by Huchtmeier et al.(1997,1999) and Cote et al.(1997).
Our updated version of the KKT sample contains a total of 303 galaxies.
  For the galaxies with measured HI flux, S (Jansky km/s), corrected for
beam-filling effect
we calculated the mass of its neutral hydrogen

	 $$    \lg(M_{HI}/M_{\sun})= \lg S + 2 \lg D + 5.37, $$
where $D$ is the distance to a galaxy in Mpc. For 188 galaxies we used
individual photometric distance estimates. For 47 galaxies located within 
the known groups we adopted the median distance of each group, and for
 83 galaxies their distances were determined via radial velocity:
$D= V_0/H$. The local value of the Hubble parameter was taken to be $H$ = 70 km/s/Mpc.
The radial velocity of galaxies, $V_0$, was corrected for the solar motion with
respect to the Local Group centroid towards ${l_A= 93\degr, b_A= -4\degr}$ with the
$V_A= 316$ km/s (Karachentsev \& Makarov, 1996). Calculating the total blue
luminosity of the galaxy, $L/L_{\sun}$, and its standard linear diameter, $A_{25}$
(in Kpc), we took into account corrections for the Galactic extinction and
the galaxy inclination in the manner adopted in PGC. For heavy obscured objects
we used extinction values from the IR sky map by Schlegel et al.(1998).
To estimate the "total" mass enclosed inside the standard angular
diameter, $a_{25}$, (in arcmin) we used the relation

	 $$ \lg(M_{25}/M_{\sun}) = 2\lg V_m + \lg a_{25} +lgD + 4.52, $$
where $V_m$ means the galaxy rotation amplitude (in km/s) corrected for
inclination and turbulence. For dwarf galaxies inclination values are rather uncertain
as they are normally derived from uncertain axial ratios. The transformation of the 
galaxy HI line width
$W_{50}$ or $W_{20}$ into $V_m$ was done following the model of corrections for
turbulent motions by Tully (1988) with parameters $\sigma_t = 19$ km/s and $\sigma_z =
12$ km/s.  Due to the uncertainty of correction for inclination of galaxies
having axial ratio $b/a > 0.8$ we did not determine the total mass of
such galaxies.

  The resulting list of basic parameters for the considered galaxies is
given in the Appendix. Its columns contain: (1) the galaxy name (And III,
Cas 1,etc.) or number in different catalogues and lists: NGC (N), IC, UGC (U),
UGCA (UA), PGC (P), DDO, KK (Karachentseva \& Karachentsev, 1998), (2) its
rough coordinates on (1950.0), (3) blue total magnitude, (4) corrected
radial velocity, (5) morphological type, (6) distance in Mpc with indication
of its origion: P - photometric, M - from membership, H - from the Hubble low,
(7) tidal index, (8) logarithm of corrected linear diameter in Kpc, (9) logarithm
of luminosity in solar units, (10) logarithm of rotational velocity in $km\,s^{-1}$  
corrected for inclination, (11) logarithm of "total" mass-to-luminosity ratio
in solar units, (12) logarithm of HI mass-to-luminosity ratio in solar units,
(13) logarithm of fractional HI mass, (14) mean optical surface brightness in
$mag/square\,arcsec$. In total the Appendix contains data on 275 galaxies with
$V_0 < 500\, km\,s^{-1}$, and 28 dwarf spheroidal galaxies without $V_0$ but being
highly probable members of the known nearby groups. We add to the list also
15 galaxies with $500 < V_0 < 1000$ km/s having distance estimates within
10 Mpc. References to sources of data on the presented galaxies are
accessable from the e-mail address: dim@sao.ru.

\normalsize   
\begin{table*}
\caption{Coefficients of linear regression, $ y = k x + c$, for the global
	    parameters of nearby galaxies}
\begin{tabular}{llrrrrrr} \hline
\multicolumn{1}{c}{$y$}&
\multicolumn{1}{c}{$x$}&
\multicolumn{1}{c}{$N$}&
\multicolumn{2}{c}{$r(xy)$}&
\multicolumn{1}{c}{$\sigma(y)$}&
\multicolumn{1}{c}{$k$}&
\multicolumn{1}{c}{$c$}\\ \hline

  $\lg(A_{25})$& $\lg(V_m)$& 194&    78&   [74]&      0.26  &  $0.99\pm.06$&  $-0.95\pm.09$ \\
  $\lg(L)$& $\lg(V_m)$& 195 &   87&   [83] &     0.48&    $2.55\pm.10$   & $4.40\pm.17$ \\
  $\lg(A_{25})$& $\lg(L)$& 318&    92 &  [91] &     0.18&    $0.35\pm.01$ & $-2.32\pm.07$ \\
  $\lg(V_m)   $& $lg(L)$& 195 &   87 &  [83] &     0.16 &   $0.30\pm.01$ & $-0.92\pm.10$ \\
  & & & & & & &       \\
  $\lg(M_H/L) $& $\lg(L)$& 255&   $-50$&  [$-$51]& 0.46&   $-0.25\pm.03$&   $1.80\pm.23$ \\
  $\lg(M_{25}/L)$& $\lg(L)$&   194 &   $-$2 & [$-$12] &     0.35 & $-0.01\pm.03$ &   $0.47\pm.22$ \\
  $\lg(M_H/M_{25}$)&$\lg(L)$&    191&   $-57$ & [$-$39]&      0.42&   $-0.30\pm.03$&   $1.80\pm.27$ \\
  $\lg(\Sigma_{HI})$& $\lg(L)$& 255 &  $-$9 &  [$-$2]&      0.45  & $-0.04\pm.03$ &   $7.2\pm.2$\\
   & & & & & & & \\
  $\lg(M_H/L)$& $\lg(A_{25})$&   255 & $-$35&[$-$31] &     0.50 & $-0.42\pm.07$ &   $-0.1\pm.1$ \\
  $\lg(M_{25}/L)$& $\lg(A_{25})$&   194 &   12 & [4] &     0.35 & $0.10\pm.06$ &   $0.35\pm.04$ \\
  $\lg(M_H/M_{25})$& $\lg(A_{25})$&   191 &$-$49 &[$-$34] &     0.45 & $-0.60\pm.08$ &   $-0.4\pm.1$ \\
  $\lg(\Sigma_{HI})$& $\lg(A_{25})$&   255 &$-$21 &[$-$22]&     0.44 & $-0.22\pm.06$ &   $7.0\pm.1$ \\
   & & & & & & & \\
  $\lg(M_H/L)  $& $\lg(V_m)$& 191&   $-$41&  [$-$32] &     0.46 & $-0.64\pm.10$ &$0.7\pm.2$\\
  $\lg(M_{25}/L) $& $\lg(V_m)$& 194&    43 &  [38] &     0.32 & $0.46\pm.07$& $-0.31\pm.11$\\
  $\lg(M_H/M_{25})$& $\lg(V_m)$& 191&   $-$71 & [$-$64]&      0.36&  $-1.10\pm.08$& $0.98\pm.13$\\
  $\lg(\Sigma_{HI}) $& $\lg(V_m)$ &191&   $-$6 &   [12] &     0.41 & $-0.08\pm.09$& $7.00\pm.15$\\
   & & & & & & & \\
  $\lg(M_H/L)$& $\Sigma_B$& 255&    54 &    -- &      0.44&  $0.30\pm.03$& $-7.4\pm.7$\\
  $\lg(M_{25}/L) $& $\Sigma_B$ &194 &   33& -- &      0.33&  $0.12\pm.02$& $-2.5\pm.6$\\
  $\lg(M_H/M_{25})$& $\Sigma_B$ &191&    38 & --&       0.47&  $0.21\pm.04$& $-5.8\pm.9$\\
& & & & & & & \\
  $\lg(M_H/L)$ &  Type &   255&    60 &  [56] &     0.42&  $0.10\pm.01$& $-1.12\pm.07$ \\
  $\lg(M_{25}/L)$ &  Type &   194 &  $-$10 &  [9]&  0.35&  $-0.01\pm.01$& $0.51\pm.07$  \\
  $\lg(M_H/M_{25})$&  Type &   191&    71 &  [63]&   0.36&  $0.12\pm.01$&$-1.76\pm.08$  \\
 & & & & & & & \\
  $\lg(M_H/L)$& $\Theta$& 255 &  $-$11&    --&       0.52&  $-0.04\pm.02$& $-0.32\pm.03$ \\
  $\lg(M_{25}/L)$& $\Theta$& 194& $-$4 &    -- &      0.35& $-0.01\pm.02$& $0.42\pm.03$ \\
  & & & & & & &  \\
  $\lg(M_H)$&   $\lg(V_m\cdot A_{25})$& 191& 89 &    -- &      0.36&  $0.99\pm.04$& $5.9\pm.1$ \\
\hline
\end{tabular}
\end{table*}
\normalsize 

\section{Correlation statistics for basic variables}
  In our sample about 95\% of the galaxies are of irregular and spiral type.
Between their global parameters $V_m$, $A_{25},\; L_{B},\; M_{HI}$ rather close correlations
can be seen, indicating important structural and dynamic properties
of dwarf and giant galaxies. In Table 1 we present numerical parameters
for the linear regression,  $y = k\cdot x + c$, where  the variables $x$  and $y$
correspond to logarithms of different integral characteristics of the galaxies.
The table columns contain: $N$ --- the number of galaxies in the subsample;
$r(x,y)$ --- the correlation coefficient in per cent, its value from  Huchtmeier \&
Richter (1988) is given in brackets; $\sigma(y)$ --- the standard deviation with
respect to the regression line; $k$ and $c$ --- the regression parameters with their
standard deviations.

  Two upper lines in the table describe the Tully-Fisher relation for the
galaxies with $V_0 < 500$ km/s. A comparison of  $r(x,y)$ values for our
sample with the earlier data from HR shows  good agreement. 
The scatter of galaxies on the Tully-Fisher diagrams appears to be lower
when new photometric data and photometric distance estimates  are
used instead of kinematic ones, $V_0/H$. It should be noted that the linear
diameter of a galaxy and its rotational velocity follow a linear relation with
$k= 0.99\pm0.06$ in the whole range of diameters: from 1 Kpc to 40 Kpc. The
same property was found also for thin disk-like galaxies viewed edge-on
(Karachentsev et al., 1999). Apparently, the linear relation $A_{25}\propto V_m$
has a fundamental kinematic significance, reflecting conditions of formation
and equalibrium of gaseous disks of galaxies.

  Due to the tight correlations between luminosity, linear diameter and
rotational velocity of the galaxies each of these parameters may be considered
as a suitable argument to distinguish between giant, normal and dwarf objects.
However, below we give preference to $V_m$ as a variable which is independent
of distance determination errors.
%Fig. 1
\begin{figure}
\caption{The relationship between the hydrogen mass-to-luminosity ratio and
       the rotational velocity. 
       The solid line shows the least-squares
       regression. Some galaxies with extreme parameters, like NGC 205,
       DDO 154, UGCA 292, are indicated with their name in the figure.
       The quantities "r" and "k" in a corner correspond to the regression
       parameters in the columns (4) and (7) of Table 1.}
\end{figure} 

  Figure 1 presents the distribution of the LV galaxies according to their
rotational velocity and hydrogen mass-to-luminosity ratio. Here  both
variables are independent of the galaxy distance. These data confirm the well
-known fact ( HR, McGaugh \& de Blok, 1997 ) that the amount of hydrogen mass
per unit of luminosity increases from giant spirals towards dwarfs. For some
dwarf systems (K~90, DDO~154, and UGCA~292) their $M_{HI}/L$ ratio
reaches the maximum value, $\sim5 M_{\sun}/L_{\sun}$.
The distribution of galaxies according to the "total" mass-to-luminosity ratio
and rotational velocity is given in Fig.2. Unlike $M_{HI}/L,$ the $M_{25}/L$
ratio  tends to decrease from giant spirals to dwarf galaxies.
The same result was derived by HR (line 14 in Table 1) and Broeils \& Rhee (1997).
It should be noted, however, that $ M_{25}/L$ is practically independent of
the galaxy luminosity (line 6 in Table 1). Moreover, some authors
(HR, Salpeter \& Hoffman, 1996) point even to a small increase in $M_{25}/L$
towards dwarf galaxies, which gives grounds to assume a growth of relative amount
of Dark Matter towards dwarf galaxies. But the origin of this difference may
simply be caused by the statistical nature of the relations: $M_{25}/L\propto V_m$
and $M_{25}/L\propto L$, when mesurement errors of the observables have
different influence on the correlation coefficients. 

%Fig.2 
\begin{figure} 
\caption{The "total" mass-to-luminosity ratio versus the rotational velocity.
       The least-squares regression parameters, r and k, from Table 1 
       are shown in a
       corner.}
\end{figure} 

As it is seen in
Fig.2, the value of $M_{25}/L$ for the considered galaxies occupies a range from
0.2 to 16 $M_{\sun}/L_{\sun}$  with a median of 3.0 $M_{\sun}/L_{\sun}$. The minimum
"total" mass-to-luminosity ratios are characteristic of galaxies having a high
surface brightness with signs of active star formation (NGC~1569, NGC~5253).
The maximum $M_{25}/L$ ratios are inherent in galaxies of low surface brightness
like KK~210, PGC~18370, K~15 and K~90.

  Staveley-Smith \& Davies (1988) and Huchtmeier \& Richter (1988) noted
that the hydrogen mass-to-"total" mass
ratio increases from giant  towards dwarf systems. This known
effect is also well seen in Fig.3 for the sample of nearby galaxies. 
%Fig.3 
\begin{figure} 
\caption{The fractional HI mass shown as a function of rotational velocity. 
       The solid line shows the least-squares
       regression. Some galaxies with extreme parameters, like NGC 205,
       DDO 154, UGCA 292, are indicated with their name in the figure.
       The quantities "r" and "k" in a corner correspond to the regression
       parameters in the columns (4) and (7) of Table 1.}
\end{figure} 
This
relation has a clearer shape, when the rotational velocity is used as the
argument instead of luminosity or linear diameter (see lines 7,11, and 15 in
Table 1). The median value of $M_{HI}/M_{25}$ for the LV galaxies is about
0.25. Several dwarf systems like UGC~7949, K~215, and UGCA~292 have
$M_{HI}/M_{25}$ in the range of [1 -- 3], which suggests that the true
total mass of some dwarf galaxies exceeds their mass within $R_{25}$ at least
by (2 -- 3) times in accordance with Broeils (1992). An example of such systems
is DDO~154 (Carignan \& Beaulieu, 1989). But the three objects mentioned above
seem to be even more HI-extended and unusual, deserving a detailed kinematic
study in the HI line.

%Fig. 4
\begin{figure} 
\caption{The HI surface density within the standard linear diameter (in $M_0$ per
       Kpc$^2$) as a function of rotational velocity. 
       The solid line shows the least-squares
       regression. Some galaxies with extreme parameters, like NGC 205,
        are indicated with their name in the figure.
       The quantities "r" and "k" in a corner correspond to the regression
       parameters in the columns (4) and (7) of Table 1.}
\end{figure}  
In Fig.4 a plot of the HI surface density (in $M_{\sun}$ per Kpc$^2$) versus $V_m$
is displayed. The least-squares fit of these data, shown by the solid line,
yields a slight decrease in $\Sigma_{HI}$ towards giant galaxies unsignificant
at the 1-sigma level.

\section{Effects of optical surface brightness and type}
  Above we briefly described the relations of basic global parameters of
nearby galaxies with their blue luminosity or rotational velocity. Bothun
et al., (1997) and some other authors argued that galaxies of
high and very low surface brightness have essentially different conditions
for star formation. This must lead to a difference in their
HI content and global structure. In Fig.5 the hydrogen mass-to-luminosity
ratio for the LV galaxies is plotted versus the mean blue surface brightness.
These data show that the relative content of hydrogen drops apparently
with increasing surface brightness. It may be caused by evolutionary
transformation  of the gaseous  component of the galaxy into its stellar
population (McGaugh \& de Blok, 1997).

%Fig.5 
\begin{figure} 
\caption{The HI mass-to-blue luminosity ratio versus the mean blue surface
       brightness (in mag/arcsec$^2$). 
       The solid line shows the least-squares
       regression. Some galaxies with extreme parameters, like NGC 205,
        are indicated with their name in the figure.
       The quantities "r" and "k" in a corner correspond to the regression
       parameters in the columns (4) and (7) of Table 1.}
\end{figure} 
%Fig. 6
\begin{figure} 
\caption{The total mass-to-luminosity ratio as a function of optical surface
       brightness. 
       The solid line shows the least-squares
       regression. Some galaxies with extreme parameters, like NGC 205,
       are indicated with their name in the figure.
       The quantities "r" and "k" in a corner correspond to the regression
       parameters in the columns (4) and (7) of Table 1.}
\end{figure} 
On the average the hydrogen mass-to-luminosity ratio varies by a factor of $\sim30$
with a maximum variation of the mean surface brightness by 5 magnitudes.
A similar diagram for the total mass-to-luminosity ratio (Fig.6) reveals only
a slight correlation with the optical surface brightness of the galaxy.

  According to Roberts \& Haynes (1994) the mean ratios, $M_{HI}/L$ and $M_{HI}/M_{25},$
increase smoothly from early morphological types towards late ones, but the
ratio $<M_{25}/L\mid T>$ appears to be approximately constant. In the considered volume-limited
sample the early type galaxies are represented only in  small numbers. Nevertheless,
the regression coefficients in lines 20--22 of Table 1 agree well with
the data from Roberts \& Haynes derived for the Local Supercluster sample.

%Fig.7
\begin{figure} 
\caption{The total mass-to-luminosity ratio against the morphological type.
       The solid line shows the least-squares regression, the dashed
       line is the relation for a purely stellar population of galaxies
       caused by their evolution according to Hoffman et al.,1996.}
\end{figure} 
In Fig.7 a plot of the $M_{25}/L$ versus morphological type is shown for our
sample, where the straight line represents the linear regression. The dashed
line refers to the variation of $M_{25}/L$ along the morphological sequence
for a purely stellar population of galaxies caused by their evolution (Larson
\& Tinsley, 1978, Hoffman et al., 1996).

\section{Environment effects}
  Many publications  describe a segregation of galaxies
by morphological types and HI content as dependent on the number density of
surrounding galaxies. The most distinct deficiency of HI-rich galaxies is
seen in the central parts of clusters (Dressler, 1984, Giovanelli \& Haynes,
1991). But the data on segregation outside galaxy clusters look rather
controversal.

  To describe the local mass density around a galaxy "i" we introduced the
so-called "tidal index" or isolation index
 (Karachentsev \& Makarov, 1998):

      $$\Theta_i = \max\{log(M_k/D^3_{ik})\} + C, \;     i = 1, 2,....N ,$$
where $M_k$ is the total mass of any neighbouring galaxy separated from the
considered galaxy by a distance of $D_{ik}$. For every galaxy "i" we found its
"main disturber", producing the maximum density enhancement, $\Delta\rho\sim
M_k/D^3_{ik}$. The value of constant $C$ is choosen so that $\Theta=0$
when the Keplerian cyclic period of the galaxy with respect to its main
disturber equals the cosmic Hubble time, 1/H. Therefore, galaxies
with $\Theta < 0$ may be considered as well isolated objects.

%Fig. 8 
\begin{figure} 
\caption{The plot of HI mass-to-luminosity ratio versus the "tidal index"
       defined above and proportional to the local mass density. 
       The solid line shows the least-squares
       regression. Some galaxies with extreme parameters, like NGC 205,
        are indicated with their name in the figure.
       The quantities "r" and "k" in a corner correspond to the regression
       parameters in the columns (4) and (7) of Table 1.}
\end{figure} 
  The distribution of the LV galaxies in $M_{HI}/L$ versus index $\Theta$ is
plotted in Fig.8. As it can be seen from this diagram, in the whole range of
local densities, $\Delta\lg\rho_k = 6$, the mean hydrogen mass-to-luminosity
ratio remains approximately constant, showing an unsignificant segregation
effect from environment. It should be, however, noted that many spheroidal dwarf
galaxies, whose HI-fluxes lie below a threshold detection, are not represented
in Fig.8, which produces an effect of observational selection. Another global
parameter of galaxies, $M_{25}/L$, seems also to have the same mean value as for
very isolated galaxies as well as for tight companions of massive galaxies (see
line 24 in Table 1).

\section{HI mass and angular momentum}
  For the disks of gas-rich galaxies, where the active star formation is
going on, the amount of hydrogen mass and angular momentum, proportional to
($V_m\cdot A_{25})$, must follow a linear regression: $M_{HI}\propto V_m A_{25}$ 
(Zasov, 1974).
For 134 spiral galaxies Zasov \& Rubtzova (1989) derived a relation
$\lg M_{HI}\propto(1.3\pm0.1)\lg(V_m\cdot A_{25})$. Considering a sample of 535 
thin (buldgeless)   
edge-on galaxies Karachentsev et al. (1999) found that the slope of this
relation is close to unity, $k = 1.08\pm0.03$, in the range of angular
momentum covering two orders.

%Fig.9 
\begin{figure} 
\caption{The HI mass as a function of angular momentum represented by a
       product of rotational velocity of the galaxy on its standard linear
       diameter. 
       The solid line shows the least-squares
       regression. 
       The quantities "r" and "k" in a corner correspond to the regression
       parameters in the columns (4) and (7) of Table 1.}
\end{figure} 
Fig.9 shows the relation between hydrogen mass and angular momentum
for the LV galaxies. Because of the presence of a large number of small nearby
galaxies, the range of their angular momentum extends over three orders.
In this wide range the relation has a slope $k = 0.99\pm0.04$, correlation
coefficient  $r = 0.89$, and the standard deviation $\sigma(\lg M_{HI})= 0.36$.
Therefore, we may consider that the gaseous
disks of giant spiral galaxies as well as dwarf irregular ones are
situated, apparently, near the threshold of gravitational instability,
favouring active star formation in them.

\section{Conclusion}
  To study relations between global optical and HI parameters of galaxies
we used a nearly complete sample of nearby galaxies with corrected radial
velocities, $V_0 < 500$ km/s (the KKT sample). Comparing with the previous
investigation undertaken by Huchtmeier \& Richter (1988), the initial KKT
sample was updated by new nearby objects and also by new more homogeneous
data on galaxy distances, apparent magnitudes and other observational
parameters. A survey of basic relations between the global characteristics
of nearby galaxies is presented in Table 1. We note the  following
properties among them:

  1) In the whole range of diameters from 1 Kpc to 40 Kpc the LV galaxies
follow the Tully-Fisher relation, $\lg A_{25}\propto(0.99\pm0.06)\lg V_m$,
whose linear character may have a deep evolutionary sense.

  2) The hydrogen mass-to-luminosity and hydrogen mass-to-"total" mass ratios
increase systematically from giant galaxies to dwarfs. The median value of
$M_{HI}/M_{25}$ is 0.25, however, some very nearby dwarf galaxies, such as
UGCA~292, having received no detailed study as yet, have $M_{HI}/M_{25} = 1 - 3.$

  3) For the LV galaxies their "total" mass (inside the standard optical
diameter)-to-the total blue luminosity ratios are concentrated in the range
of 0.2 to 16 $M_{\sun}/L_{\sun}$ with the median  3.0 $M_{\sun}/L_{\sun}$.
This ratio tends to decrease slightly from  giant
galaxies to dwarfs, which needs an explanation within the common
idea of presence of large amount of Dark Matter in dwarf irregular systems.

  4) Our sample galaxies differ in their mean surface brightness almost by a factor
of 100. The hydrogen mass-to-luminosity ratio for them shows a clear-cut
increase towards the low surface brightness objects. A similar but less
distinct relation is also seen for the total mass-to-luminosity ratios.
Both correlations may be caused by star formation processes going on in the
galaxies.

  5) To quantitatively estimate the local mass density around each LV galaxy,
produced by its neighbours, we used the so-called "tidal index". Despite the
differences of about six orders in the local densities, the ratios $M_{HI}/L$ and
$M_{25}/L$ do not show pronounced correlation with the tidal index similar to
the HI-deficit or morphological segregation seen in rich galaxy clusters.

  6) The amount of hydrogen mass in a galaxy and its angular momentum follow
a nearly linear relation, $\lg M_{HI}\propto(0.99\pm0.04)\lg(V_m\cdot A_{25})$ in the range
of angular momentum which exceeds three orders. According to Zasov (1974) it
means that the gaseous disks of giant, normal, and dwarf galaxies are situated
near the threshould of gravitational instability favouring star formation
in them.

\acknowledgements{This work is supported by INTAS-RFBR grant 95-IN-RU-1390
and DFG grant KS~9112.}

{}

\begin{thebibliography}{}

\bibitem{} Bothun G., Impey C., McGaugh S., 1997, PASP 109. 745
\bibitem{} Broeils A.H., 1992, Dark and Visible Matter in Spiral Galaxies, Dissertation,
		 Groningen
\bibitem{} Broeils A.H., Rhee M.H., 1997, A\&A, 324. 877
\bibitem{} Carignan C., Beaulieu S., 1989, ApJ 347. 760
\bibitem{} Giovanelli R.G., Haynes M.P., 1991, ARA\&A 29. 499
\bibitem{} Cote S., Freeman K.C., Carignan C., Quinn P.J., 1997, AJ 114. 1313
\bibitem{} Dressler A., 1984, ARA\&A, 22. 185
\bibitem{} Hoffman G.L., Salpeter E.E., Fathat B., Roos T., Williams H., Helou G., 1996,
	    ApJS 105. 269
\bibitem{} Huchtmeier W.K., Karachentsev I.D., Karachentseva V.E., 1997, A\&A 322. 375
\bibitem{} Huchtmeier W.K., Karachentsev I.D., 1999, A\&A, in prepapation
\bibitem{} Huchtmeier W.K., Richter O.G., 1988, A\&A 203. 237 (HR)
\bibitem{} Karachentsev I.D., 1994, Astron. \& Astrophys. Transcations, 6, 3
\bibitem{} Karachentsev I.D., Tikhonov N.A., 1994, A\&A, 286, 718
\bibitem{} Karachentsev I.D., Karachentseva V.E., Kudrya, Yu.N., 1999, Letters to Astron.
		 Zh., 25, 3
\bibitem{} Karachentsev I.D., Makarov D.I., 1996, AJ 111. 535
\bibitem{} Karachentsev I.D., Makarov D.I., 1998, in Procccedings of IAU Symp. No 186,
		Kyoto, 109
\bibitem{} Karachentseva V.E., Karachentsev I.D., 1998, A\&AS, 127, 409
\bibitem{} Kraan-Korteweg R.C., Tammann G.A., 1979, Astron. Nachr. 300. 181 (KKT)
\bibitem{} Larson R.B., Tinsley B.M., 1978, ApJ 219. 46
\bibitem{} McGaugh S.S., de Blok W.J.G., 1997, ApJ 481. 689
\bibitem{} Paturel G., Fouque P., Bottinelli L., Gouguenheim L., 1992, Catalogue of
	    Principal Galaxies, Lyon (PGC)
\bibitem{} Roberts M.S., 1969, AJ, 74, 859
\bibitem{} Roberts M.S., Haynes M.P., 1994, ARA\&A 32. 115
\bibitem{} Salpeter E.E., Hoffman G.L., 1996, ApJ 465. 595
\bibitem{} Schlegel D.J., Finkbeiner D.P., Davis M., 1998, ApJ, 500, 525
\bibitem{} Staveley-Smith L., Davies R.D., 1988, MNRAS 231. 833
\bibitem{} Tully R.B., 1988, Nearby Galaxy Catalog, Cambridge Univ. Press
\bibitem{} Tully R.B., Fisher J.R., 1977, A\&A 54. 661
\bibitem{} Zasov A.V., 1974, AZh 51. 1225
\bibitem{} Zasov A.V., Rubtzova T.V., 1989, Letters to Astron. Zh. 15. 118
\end{thebibliography}
\end{document}